\documentclass[aps,prd,twocolumn,floatfix,nofootinbib,showpacs,superscriptaddress,tightenlines]{revtex4}

\usepackage{amsmath}
\usepackage{lipsum}
\usepackage{amssymb}
\usepackage{amsthm}
\usepackage{bbold}
\usepackage{dcolumn}
\usepackage{epsfig}
\usepackage{graphics}
\usepackage{graphicx}
\usepackage{color}
\usepackage{xspace}
\usepackage{cancel}

\newcommand{\MeV}{\text{MeV}} 
\newcommand{\GeV}{\text{GeV}} 

\newcommand{\bea}{\begin{eqnarray}}
\newcommand{\eea}{\end{eqnarray}}
\newcommand{\nn}{\nonumber}

\newcommand{\eqn}[1]{Eq.~(\ref{#1})}
\newcommand{\fig}[1]{Fig.~\ref{#1}}
\newcommand{\tab}[1]{Table~\ref{#1}}
\newcommand{\sect}[1]{Section~\ref{#1}}

\newcommand{\df}[1]{\hspace{-0.5em}\ensuremath{\frac{\mathrm{d}^{4}#1}{(2\pi)^{4}}}\,}
\newcommand{\dx}[1]{\hspace{-0.5em}\ensuremath{\mathrm{d}#1}\,}

\newcommand{\Tr}{\text{Tr}}

\begin{document}

\title{Heavy quarkonia in a contact interaction and an algebraic model: mass spectrum,
decay constants, charge radii and elastic and transition form
factors}

\author{Kh\'epani Raya}
\affiliation{ Instituto de F\'{i}sica y Matem\'aticas, Universidad
Michoacana de San Nicol\'as de Hidalgo, Edificio C-3, Ciudad
Universitaria, Morelia, Michoac\'an 58040, M\'exico}
\author{Marco A. Bedolla}
\affiliation{ Instituto de F\'{i}sica y Matem\'aticas, Universidad
Michoacana de San Nicol\'as de Hidalgo, Edificio C-3, Ciudad
Universitaria, Morelia, Michoac\'an 58040, M\'exico} \affiliation{
Istituto Nazionale di Fisica Nucleare (INFN), Sezione di Genova,
via Dodecaneso 33, 16146 Genova, Italy}
\author{J.J. Cobos-Mart\'{\i}nez}
\affiliation{ Laborat\'orio de F\'{i}sica Te\'orica e
Computacional, Universidade Cruzeiro do Sul, 01506-000 S\~ao
Paulo, Brasil} \affiliation{C\'atedra CONACyT, Departamento de
F\'isica, Centro de Investigaci\'on y de Estudios Avanzados del
Instituto Polit\'ecnico Nacional, Apartado Postal 14-740, 07000,
Ciudad de M\'exico, M\'exico}
\author{Adnan Bashir}
\affiliation{ Instituto de F\'{i}sica y Matem\'aticas, Universidad
Michoacana de San Nicol\'as de Hidalgo, Edificio C-3, Ciudad
Universitaria, Morelia, Michoac\'an 58040, M\'exico}

\date{\today}

\begin{abstract}

For the flavor-singlet heavy quark system of bottomonia, we
compute the masses of the ground state mesons in four different
channels, namely, pseudo-scalar ($\eta_{b}(1S)$), vector
($\Upsilon(1S)$), scalar ($\chi_{b_0}(1P)$) and axial vector
($\chi_{b_{1}}(1P)$). We also calculate the weak decay constants
of the $\eta_{b}(1S)$ and $\Upsilon(1S)$ as well as the charge
radius of $\eta_{b}(1S)$. It complements our previous study of the
corresponding charmonia systems: $\eta_c(1S)$, $J/\Psi(1S)$,
$\chi_{c_0}(1P)$) and ($\chi_{c_{1}}(1P)$). The unified formalism
for this analysis is provided by a symmetry-preserving
Schwinger-Dyson equations treatment of a vector$\times$vector
contact interaction. Whenever a comparison is possible, our
results are in fairly good agreement with experimental data and
model calculations based upon Schwinger-Dyson and Bethe-Salpeter
equations involving sophisticated interaction kernels. Within the
same framework, we also report the elastic and transition form
factors to two photons for the pseudo-scalar channels
$\eta_{c}(1S)$ and $\eta_{b}(1S)$ in addition to the elastic form
factors for the vector mesons $J/\Psi$ and $\Upsilon$ for a wide
range of photon momentum transfer squared ($Q^2$). For
$\eta_{c}(1S)$ and $\eta_{b}(1S)$, we also provide predictions of
an algebraic model which correlates remarkably well between the
known infrared and ultraviolet limits of these form factors.


\end{abstract}


\maketitle

\date{\today}

\section{\label{sec:intro}Introduction}

        The discovery of heavy and narrow $J/\psi$ resonance in
1974~\cite{Aubert:1974js, Augustin:1974xw} was followed by an even
heavier and narrower $\Upsilon$ resonance in
1977~\cite{Herb:1977ek}. These systems are characterized by two
diametrically opposite energy scales: the hard scale of the heavy
constituent quark masses and the soft scale of the relative
momenta between them. They are bound systems yet their diminishing
size with increasing mass probes coupling which tends to approach
asymptotic freedom. Schwinger-Dyson equations (SDEs) of quantum
chromodynamics (QCD) provide a natural means to study these
systems; their derivation makes no assumption about the strength
of the interaction involved. Therefore, they are ideally suited to
study systems or phenomena which probe different energy scales
simultaneously.

Earliest studies of heavy quarkonia through SDEs can be traced
back to Ref.~\cite{Jain:1993qh}. With refined truncations of these
equations and/or increased numerical complexity, these systems
have also been investigated in
Refs.~\cite{Krassnigg:2004if,Bhagwat:2004hn,
Bhagwat:2006xi,Maris:2005tt,Souchlas:2010zz,Blank:2011ha,Rojas:2014aka,Fischer:2014cfa,Ding:2015rkn,Raya:2016yuj}.
Predictions for states with exotic quantum numbers were made in
Refs.~\cite{Maris:2006ea, Krassnigg:2009zh}.

The extension of this program to the complicated exotic and
baryonic states, decay rates and form factors is considerably
non-trivial. Brute force numerical evaluation stops short of
exploring the large momentum transfer region of form factors and,
at times, is unable to make full comparison with already available
experimental data. This shortcoming has been exposed in the
calculation of elastic form factors (EFFs), see
Ref.~\cite{Maris:2000sk} for the pion EFF, as well as transition
form factors (TFF), (see {\em e.g.}, Ref.~\cite{Chen:2016bpj} for
the two photon TFF of $\eta_c$). However, an artful
parameterization of the Bethe-Salpeter amplitudes (BSAs) in terms
of Nakanishi-like perturbation theory integral
representations~\cite{Nakanishi:1963zz} allows us to reach large
space-like momentum transfer region, see for example
Refs.~\cite{Chang:2013nia,Raya:2015gva,Raya:2016yuj}.

A few years ago, an alternative to full QCD based explorations was
put forward to study pion properties assuming that quarks
interact, not via massless vector-boson exchange, but instead
through a symmetry preserving vector-vector contact interaction
(CI)~\cite{GutierrezGuerrero:2010md,Roberts:2010rn,
Chen:2012qr,Roberts:2011cf,Roberts:2011wy}. One then proceeds by
embedding this interaction in a rainbow-ladder (RL) truncation of
the SDEs. Confinement is implemented by employing a proper time
regularization scheme. This scheme systematically removes
quadratic and logarithmic divergences ensuring the axial-vector
Ward-Takahashi identity (axWTI) is satisfied. One can also
explicitly verify the low energy Goldberger-Treiman relations. A
fully consistent treatment of the CI model is simple to implement
and can help us provide useful results which can be compared and
contrasted with full QCD calculation and experiment.

This interaction is capable of providing a good description of the
masses of meson and baryon ground and excited-states for light
quarks~\cite{GutierrezGuerrero:2010md,Roberts:2010rn,
Chen:2012qr,Roberts:2011cf}. The results derived from the CI model
are quantitatively comparable to those obtained using
sophisticated QCD model
interactions~\cite{Bashir:2012fs,Eichmann:2008ae,Maris:2006ea,Cloet:2007pi}.
Interestingly and importantly, this simple CI produces a
parity-partner for each ground-state that is always more massive
than its first radial excitation so that, in the nucleon channel,
e.g., the first $J^P = {1}/{2}^-$ state lies above the second $J^P
= {1}/{2}^+$ state~\cite{Chen:2012qr,Lu:2017cln}.

Considering the discussion so far as a satisfactory justification,
in Ref.~\cite{Bedolla:2015mpa}, we extended this interaction model
to the sector of heavy quarkonia, computing the mass spectrum of
charmonia as well as the decay constants of the pseudoscalar and
vector meson channels. A subsequent
publication~\cite{Bedolla:2016jjh} applies this model to the
computation of the EFF and the TFF of the $\eta_c$. Low momentum
limit of the form factors allows us to extract charge radii which
compare well with the earlier SDE computations and the lattice
results. A symmetry-preserving CI has also been employed recently
to study charmed mesons in~\cite{Serna:2017nlr}.

Building on our efforts, we now extend this interaction to the
analysis of the flavor-singlet heavy quark system of bottomonia,
computing the masses of the ground state mesons in four different
channels: pseudo-scalar ($\eta_{b}(1S)$), vector ($\Upsilon(1S)$),
scalar ($\chi_{b_0}(1P)$) and axial vector ($\chi_{b_{1}}(1P)$),
as well as the {\em weak decay constants} of the $\eta_{b}(1S)$
and $\Upsilon(1S)$. We also compute the EFFs of $\eta_{b}(1S)$,
$J/\psi$ and $\Upsilon(1S)$, complementing our previous effort.
Through evaluating the slope of these plots at the zero momentum
transfer, we calculate and report their charge radii. We also
calculate the transition form factor for $\eta_{b} \rightarrow
\gamma^* \gamma$. Wherever possible, we compare our findings with
experiment and/or other similar studies. Expectedly, the form
factors are harder than we expect from the proper treatment of
full QCD with a running quark mass function. As a quick fix, one
resorts to the astutely constructed algebraic model (AM) and finds
a near perfect substitute of cumbersome QCD based calculations.

The paper is organized as follows: in~\sect{sec:sde-bse} we
present the minimum but self contained details of the SDE-BSE
approach to mesons; we also introduce the RL truncation for the CI
and the consequences it has for the interaction kernels. We end
the section by recalling the AM which mimics QCD to an adequate
extent. In~\sect{sec:massspectrum}, we tabulate our findings for
the mass spectrum of the ground state bottomonia and decay
constants of $\eta_{b}(1S)$ and $\Upsilon(1S)$. \sect{sec:eff}
details algebraic expressions and numerical results for the EFFs
of the pseudoscalar and vector mesons. It also contains a
discussion on the quark-photon vertex we put to use, consistent
with the WTI in the RL with a CI. \sect{sec:tff} is devoted to the
results and analysis of the transition form factors for the
processes $\eta_{c(b)}\gamma^{*}\rightarrow\eta_{c(b)}$ and
$\gamma\gamma^{*}\rightarrow\eta_{c(b)}$. Finally, in
\sect{sec:conclusions}, we summarize our conclusions.

\section{\label{sec:sde-bse} SDE-BSE formalism}

We devote this section to the brief recapitulation of the SDE-BSE
formalism to study two-particle bound states and their connection
with chiral symmetry breaking. Our particular focuss will
naturally be on the CI which is employed to produce the bulk of
results reported in this article.

\subsection{\label{sec:rl-ci} The Gap Equation and the Contact Interaction}

%
The $f$-flavor dressed-quark propagator $S_{f}$ is obtained by
solving the quark
SDE~\cite{Roberts:2007jh,Holl:2006ni,Maris:2003vk,Alkofer:2000wg}
\begin{eqnarray}
 \label{eqn:quark_sde}
 &&\hspace{-0.5cm} S_{f}^{-1}(p)=i\gamma\cdot p + m_{f} + \Sigma_{f}(p) \,, \\
 \label{eqn:quark_se}
 && \hspace{-0.5cm} \Sigma_{f}(p)=
\int\df{q}
g^{2}D_{\mu\nu}(p-q)\frac{\lambda^{a}}{2}\gamma_{\mu}S_{f}(q)\Gamma^{a}_{\nu}(p,q)
\,,
\end{eqnarray}
\noindent where $g$ is the strong coupling constant, $D_{\mu\nu}$
is the dressed gluon propagator, $\Gamma^{a}_{\nu}$ is the dressed
quark-gluon vertex, $m_{f}$  is the $f$-flavor current-quark mass
and $\lambda^a$ are the usual Gell-Mann matrices.

Both $D_{\mu\nu}$ and $\Gamma^{a}_{\nu}$ satisfy their own SDEs,
which in turn are coupled to higher $n$-point functions and so on
{\it ad infinitum}. Therefore, the SDEs form an infinite set of
coupled nonlinear integral equations, requiring a truncation
scheme in order to define a tractable problem. This is achieved
once we have specified the gluon propagator and the quark-gluon
vertex. For a comprehensive recent review of the SDE-BSE formalism
and its applications to hadron physics, see for example
Refs.~\cite{Bashir:2012fs,Aznauryan:2012ba}.

The vector$\times$vector CI assumes that the force mediation among
quarks takes place not via massless vector-boson exchange but
instead through the interaction defined by:
\begin{eqnarray}
\label{eqn:contact_interaction}
g^{2}D_{\mu \nu}(k)&=&\frac{4\pi\alpha_{\text{IR}}}{m_g^2}\delta_{\mu \nu} \equiv
\frac{1}{m_{G}^{2}}\delta_{\mu\nu} \,, \\
\label{eqn:quark_gluon_vertex_rl}
\Gamma^{a}_{\mu}(p,q)&=&\frac{\lambda^{a}}{2}\gamma_{\mu} \,,
\end{eqnarray}
\noindent where $m_g=800\,\MeV$ is a gluon mass scale generated
dynamically in QCD (see for example Ref.~\cite{Boucaud:2011ug})
and $\alpha_{\text{IR}}$ is a parameter which specifies the
enhanced interaction strength in the infrared
(IR)~\cite{Binosi:2016nme,Deur:2016tte}.


Equations~(\ref{eqn:contact_interaction}) and
(\ref{eqn:quark_gluon_vertex_rl}) specify the kernel in the quark
SDE, \eqn{eqn:quark_sde}. In this truncation scheme, the general
solution of the $f$-flavored dressed-quark propagator is immensely
simplified~\cite{GutierrezGuerrero:2010md,
Roberts:2010rn,Chen:2012qr,Roberts:2011cf,Roberts:2011wy,Bedolla:2015mpa,
Bedolla:2016jjh}:
\begin{equation}
\label{eqn:quark_inverse_contact} S_{f}^{-1}(p)= i \gamma\cdot p +
M_{f} \,.
\end{equation}
\noindent As the interaction,
Eqs.~(\ref{eqn:contact_interaction},\ref{eqn:quark_gluon_vertex_rl}),
is momentum independent, the fermion mass $M_f$ follows suit. This
flavor-dependent constant mass is obtained by solving
\begin{equation}
  \label{eqn:const_mass} M_{f} = m_{f} +
  \frac{16M_{f}}{3\pi^{2}m_{G}^{2}}\int\df{q}\frac{1}{q^{2}+M_{f}^{2}} \,.
\end{equation}
Since the integral in \eqn{eqn:const_mass} is divergent, we must
adopt a regularization procedure. We employ the proper time
regularization scheme~\cite{Ebert:1996vx} to write this equation
as
\begin{eqnarray}
  \label{eqn:const_mass_reg}
 M_{f}&=& m_{f} + \frac{M_{f}^3}{3\pi^{2}m_{G}^{2}}
  \label{eqn:Ifun}
  \Gamma(-1,\tau_{\text{UV}}^{2}M_{f}^{2},\tau_{\text{IR}}^{2}M_{f}^{2}) \,,
\end{eqnarray}
\noindent where $\Gamma(a,z_{1},z_{2})$ is the generalized
incomplete Gamma function:
 \bea
 \Gamma(a,z_{1},z_{2}) =\Gamma(a,z_{1})- \Gamma(a,z_{2}) \,. \nn
 \eea
The parameters $\tau_{\text{IR}}$ and $\tau_{\text{UV}}$ are,
respectively, infrared and ultraviolet regulators. A nonzero value
for $\tau_{\text{IR}}\equiv 1/\Lambda_{\text{IR}}$ implements
confinement by ensuring the absence of quark production
thresholds~\cite{Roberts:2007ji}. Since the CI is not a
renormalizable theory, $\tau_{\text{UV}}\equiv
1/\Lambda_{\text{UV}}$ plays a dynamical role. Therefore, it sets
the scale for all dimensioned quantities. The importance of an
ultraviolet cutoff in Nambu--Jona-Lasinio type models has also
been discussed in Refs.~\cite{Farias:2005cr,Farias:2006cs}.

Note that
Eqs.~(\ref{eqn:contact_interaction},\ref{eqn:quark_gluon_vertex_rl})
not only specify the kernel in the quark SDE, \eqn{eqn:quark_sde},
but also the one in the BSE as we now make plain.

\subsection{\label{sec:BSE-ci} The BSE and the Contact Interaction}

In quantum field theory, a meson bound state in a particular
$J^{PC}$ channel is described by the
BSE~\cite{Gross:1993zj,Salpeter:1951sz,GellMann:1951rw}
\begin{equation}
\label{eqn:bse} \left[\Gamma_{H}(p;P)\right]_{tu}=
\int\df{q}K_{tu;rs}(p,q;P)\chi(q;P)_{sr} \,,
\end{equation}
\noindent where
$\chi(q;P)=S_{f}(q_{+})\Gamma_{H}(q;P)S_{g}(q_{-})$; $q_{+}=q+\eta
P$, $q_{-}=q-(1-\eta)P$; $\eta \in [0,1]$ is a momentum-sharing
parameter, $p$ ($P$) is the relative (total) momentum of the
quark-antiquark system; $S_{f (g)}$ is the $f (g)$-flavor
dressed-quark propagator, already discussed; $\Gamma_{H}(p;P)$ is
the meson Bethe-Salpeter amplitude (BSA), where $H$ specifies the
quantum numbers and flavor content of the meson; $r,s,t$, and $u$
represent color, flavor, and spinor indices; and $K(p,q;P)$ is the
quark-antiquark scattering kernel.

Eqs.~(\ref{eqn:contact_interaction},\ref{eqn:quark_gluon_vertex_rl})
define and relate the kernel of the gap equation with that of the
BSE, \eqn{eqn:bse}, through the axial-vector Ward-Takahashi
identity (axWTI)~\cite{Maris:1997hd}
\begin{equation}
  \label{eqn:axwti}
  - i P_{\mu}\Gamma_{5\mu}(k;P)=S^{-1}(k_{+})\gamma_{5} + \gamma_{5}S^{-1}(k_{-}) \,.
\end{equation}
\noindent \eqn{eqn:axwti}, which encodes the phenomenological
features of dynamical chiral symmetry breaking in QCD, relates the
axial-vector vertex, $\Gamma_{5\mu}(k;P)$, to the quark
propagator, $S(k)$. This in turn implies a relationship between
the kernel in the BSE, \eqn{eqn:bse}, and that in the quark SDE,
\eqn{eqn:quark_sde}. This relation must be preserved by any viable
truncation scheme of the SDE-BSE coupled system, thus constraining
the content of the quark-antiquark scattering kernel $K(p,q;P)$.
For the CI, \eqn{eqn:axwti} implies
\begin{equation}
  \label{eqn:bskernel_rl_contact} K(p,q;P)= -g^{2}D_{\mu\nu}(p-q)
  \left[\frac{\lambda^{a}}{2}\gamma_{\mu}\right]\otimes
  \left[\frac{\lambda^{a}}{2}\gamma_{\nu}\right] \,,
\end{equation}
\noindent where $g^{2}D_{\mu\nu}$ is given by \eqn{eqn:contact_interaction}.
Thus, the homogeneous BSE ($\eta=1$) takes the simple form
\begin{equation}
  \label{eqn:bse_contact}
  \Gamma_{H}(p;P)=-\frac{4}{3}\frac{1}{m_{G}^{2}}\int\df{q}
  \gamma_{\mu}S_{f}(q+P)\Gamma_{H}(q;P)S_{g}(q)\gamma_{\mu}.
\end{equation}
The axWTI further implies~\cite{GutierrezGuerrero:2010md,
Roberts:2010rn,Chen:2012qr,Roberts:2011cf,Roberts:2011wy,Bedolla:2015mpa,
Bedolla:2016jjh}
\begin{equation}
  \label{eqn:wticorollary}
 0=\int_{0}^{1}\dx{x}\int\df{q} \frac{\frac{1}{2}q^{2} +
    \mathfrak{M}^{2}} {\left(q^{2}+ \mathfrak{M}^{2}\right)^{2}} \,,
\end{equation}
\noindent where
 \bea
 \mathfrak{M}^{2}=M_{f}^{2}x + M_{g}^{2}(1-x)+ x(1-x)P^{2} \,. \nn
 \eea
\eqn{eqn:wticorollary} has to be faithfully maintained during the
entire calculation. It states that the axWTI is satisfied if, and
only if, the model is regularized so as to ensure there are no
quadratic or logarithmic divergences, circumstances under which a
shift in integration variables is permitted. This is an essential
requirement in order to prove
\eqn{eqn:axwti}~\cite{GutierrezGuerrero:2010md,Roberts:2010rn,Chen:2012qr,
Roberts:2011cf}.

Since the interaction kernel of \eqn{eqn:bskernel_rl_contact} does
not depend on the external relative momentum, a
symmetry-preserving regularization will return solutions which are
independent of it. Therefore, the general forms of the BSAs for
the pseudoscalar, scalar, vector, and axial-vector channels read,
respectively:
\begin{eqnarray}
  \label{eqn:psbsagral}
  \Gamma^{0^{-}}(P)&=& \gamma_{5}\left[ i E^{0^{-}}(P)
    + \frac{1}{2M}\gamma\cdot P F^{0^{-}}(P)\right] \,, \\
  \label{eqn:sbsagral}
  \Gamma^{0^{+}}(P)&=& \mathbb{1}E^{0^{+}}(P) \,, \\
  \label{eqn:vbsagral}
  \Gamma^{1^{-}}_{\mu}(P)&=&\gamma^{T}_{\mu}E^{1^{-}}(P)
  + \frac{1}{2M}\sigma_{\mu\nu}P_{\nu}F^{1^{-}}(P) \,, \\
  \label{eqn:avbsagral}
  \Gamma^{1^{+}}_{\mu}(P)&=&\gamma_{5}\left[\gamma^{T}_{\mu}E^{1^{+}}(P)
    + \frac{1}{2M}\sigma_{\mu\nu}P_{\nu}F^{1^{+}}(P)\right] \,,
\end{eqnarray}
\noindent where $M= M_f M_g/(M_f+M_g)$.
The BSE is a homogeneous equation. Thus the BSA has to be
normalized by a separate condition. In the RL approximation, this
condition is:
\begin{equation}
  \label{eqn:RLNorm}
  1=N_{c}\frac{\partial}{\partial P^{2}}\int\df{q}
  \Tr\left[\overline{\Gamma}_{H}(-Q)S(q+P)\Gamma_{H}(Q)S(q)\right] \,,
\end{equation}
\noindent evaluated at $Q=P$, where $P^{2}=-m_{H}^{2}$,
$\Gamma_{H}$ is the normalized BSA of the meson $H$, and
$\overline{\Gamma}_{H}$ is its charge-conjugated version. For the
vector and axial-vector channels, there is an additional factor of
1/3 on the right-hand side to account for all three polarizations
of a spin-1 meson.

Once the BSA has been canonically-normalized, we can extract
observables with it. The pseudoscalar and vector leptonic decay
constants, $f_{0^{-}}$ and $f_{1^{-}}$, are defined, respectively,
by
\begin{eqnarray}
  \label{eqn:psdecaydef}
  &&\hspace{-2em}P_{\mu}f_{0^{-}}=
  N_{c}\int\df{q}\Tr\left[\gamma_{5}\gamma_{\mu}
    S(q_{+})\Gamma_{0^{-}}(P)S(q_{-})\right] \,, \\
  \label{eqn:vdecaydef}
  &&\hspace{-2.75em}m_V f_{1^{-}}=
  \frac{N_{c}}{3}\int\df{q}\Tr\left[\gamma_{\mu}
    S(q_{+})\Gamma^{1^{-}}_{\mu}S(q_{-})\right] \,,
\end{eqnarray}
\noindent where $m_V$ is the mass of the vector bound state, and
the factor of 3 in the denominator corresponds to the three
polarizations of the spin-1 meson.

\subsection{The Algebraic Model}

We present here a refined version of the previously employed AM
for the heavy sector. For pseudoscalar mesons (of mass $m_P$), it
reads as:
\begin{eqnarray}
    \Gamma_P(k;P) &=& i \gamma_5 \mathcal{N}_P \frac{M}{f_P}  \int_0^1 \hspace{-0.1cm} dz \rho_\nu(z)
    \frac{M^2}{k^2+z\; \sigma_P k\cdot P + M^2}\,, \nn \\
    \rho_\nu(z)&=& \frac{1}{\sqrt{\pi}} \frac{\Gamma[\nu+3/2]}{\Gamma[\nu+1]}(1-z^2)^\nu\,,
\label{Eq:Algebraic-Model}
\end{eqnarray}
where $\mathcal{N}_P$ is the canonical normalization constant, $M$
is a mass scale (that we will relate to that of dynamical chiral
symmetry breaking) and $f_P$ is the weak decay constant; $\sigma_P
\equiv m_\pi / m_P$ is defined such that the angular dependence,
$k \cdot P$, decreases as the meson mass increases, and it
recovers the algebraic model for the pion when $m_P = m_\pi$. We
pick $\nu=1$ in accordance with our earlier study of
charmonia~\cite{Bedolla:2016jjh}.

Since there is no experimental value of $f_{\eta_b}$, we choose
$M=4.818$ GeV. It is a reasonable value since the dynamically
generated constituent-like mass generated by the CI is $M = 4.71$
GeV. Moreover, it produces the $\eta_b$ weak decay constant which
matches very well with the one reported in
Ref.~\cite{Ding:2015rkn}, based upon the most sophisticated SDE
truncation so far. It is relevant to mention that this AM produces
a finite-width delta-shaped parton distribution amplitude (PDA).
Since this is the behavior of the PDA that one expects, one would
also anticipate a decent prediction of the $\eta_b$ form factors.

We now turn to the computation of physical observables in the
following sections.

\section{\label{sec:massspectrum} Quarkonium mass spectrum}

\begin{table}[h]
\begin{center}
\begin{tabular}{c|c|c|c|c}
\hline \hline
    & \multicolumn{4}{c}{masses [GeV]}  \\
\hline
& $m_{\eta_{c}(1S)}$ & $m_{J/\Psi(1S)}$ & $m_{\chi_{c_{0}}(1P)}$ & $m_{\chi_{c_{1}}(1P)}$
\\
\hline
Experiment ~\cite{0954-3899-37-7A-075021} & 2.983 & 3.096 & 3.414 & 3.510  \\
CI~\cite{Bedolla:2015mpa} & 2.976 & 3.09 & 3.374 & 3.4 \\
JM~\cite{Jain:1993qh} & 2.821 & 3.1 & 3.605 & - \\
BK~\cite{Blank:2011ha} & 2.928 & 3.111 & 3.321 & 3.437 \\
RB~\cite{Rojas:2014aka} & 3.065 & - & - & - \\
FKW~\cite{Fischer:2014cfa} & 2.925 & 3.113 & 3.323 & 3.489 \\
DGCLR~\cite{Ding:2015rkn} & 2.98 & 3.07 & - & - \\ \hline \hline
\end{tabular}
\caption{\label{tab:mcc_all_opt} The ground-state charmonia mass
spectrum. The tabulated CI results were obtained with the best-fit
parameter set: $m_{g}=0.8\,\GeV$, $\alpha_{IR}=0.93\pi/17=0.172$,
$\Lambda_{\text{IR}}=0.24\,\GeV$ and
$\Lambda_{\text{UV}}=2.4\,\GeV$. Note that the two parameters
which differ from the light sector are $\alpha_{IR}$ and
$\Lambda_{\text{UV}}$ for the reasons explained within the text.
The current charm quark mass is $m_{c}=1.09\,\GeV$, whereas the
dynamically generated constituent-like mass is
$M_{c}=1.482\,\GeV$.
The average percentage error, with respect to experimental data,
is less than $3\%$. All subsequent calculation of charmonia
related observables in this work employs the parameters listed
this table.}
\end{center}
\end{table}

\begin{table}[h]
\begin{center}
\begin{tabular}{c|c|c|c|c}
\hline \hline
    & \multicolumn{4}{c}{masses [GeV]}  \\
\hline
& $m_{\eta_{b}(1S)}$ & $m_{\Upsilon(1S)}$ & $m_{\chi_{b_{0}}(1P)}$ & $m_{\chi_{b_{1}}(1P)}$
\\
\hline
Experiment ~\cite{0954-3899-37-7A-075021} & 9.32 (9.4) & 9.46 & 9.860 & 9.892 \\
CI [This work]& 9.345 & 9.460 & 9.594  & 9.603 \\
JM ~\cite{Jain:1993qh} & 9.322 & 9.460 & 9.860 & --- \\
BK ~\cite{Blank:2011ha} & 9.405 & 9.488 & 9.831 & 9.878 \\
FKW ~\cite{Fischer:2014cfa} & 9.414 & 9.490 & 9.815 & 9.842 \\
DGCLR~\cite{Ding:2015rkn} & 9.39 & 9.46 & - & - \\
MVRB~\cite{Mojica:2017tvh} & - & 9.552 & - & - \\
\hline \hline
\end{tabular}
\caption{\label{tab:mbb_all_opt} Ground-state bottomonia mass
spectrum. The CI results were obtained with the modified
parameters $\alpha_{IR}=0.93\pi/125=0.023$ and
$\Lambda_{\text{UV}}=6.4\,\GeV$. The current-quark mass is $m_{b}=
3.8\,\GeV$, and the dynamically generated constituentlike mass is
$M_{b}=4.7\,\GeV$.
The average percentage error, with respect to experimental data,
is again less than $3\,\%$. All bottomonia related observables in
this work have been evaluated by using the parameters listed this
table.}
\end{center}
\end{table}

In Ref.~\cite{Bedolla:2015mpa}, we extended the CI model to
charmonia. Our findings for the ground-state mass spectrum of
quark-model mesons are presented in \tab{tab:mcc_all_opt}; see
Ref.~\cite{Bedolla:2015mpa} for a discussion of these results. In
this work, we carry out an identical exercise for bottomonia.
\tab{tab:mbb_all_opt} displays results for the ground-state mass
spectrum in four different channels, obtained by solving
\eqn{eqn:bse_contact}. It can be readily inferred
from~\tab{tab:mbb_all_opt} that our results are in excellent
agreement with the experimental data and those computed more
sophisticated/complex models. That a RL truncation with a CI
describes the mass spectrum of ground-state heavy-quarkonia,
charmonia and bottomonia, so well can be easily understood because
the quark wave function renormalization is strictly one and the
mass function is momentum independent. It implies that the
heavy-quark--gluon vertex can reasonably be approximated by its
bare counterpart~\cite{Bedolla:2015mpa,Bedolla:2016jjh}.

\begin{table}[h]
\begin{center}
\begin{tabular}{c|c|c}
\hline \hline
    & \multicolumn{2}{c}{decay constants [GeV]}  \\
\hline
 & $f_{\eta_{c}}$ & $f_{J/\Psi}$
\\
Experiment & 0.238 & 0.294 \\
CI~\cite{Bedolla:2015mpa}& 0.255 & 0.206  \\
Lattice QCD & 0.279~\cite{Davies:2010ip} & 0.286~\cite{Donald:2012ga} \\
BK~\cite{Blank:2011ha} & 0.282 & 0.317 \\
KGH1~\cite{Krassnigg:2016hml} & 0.284 & 0.318 \\
KGH2~\cite{Krassnigg:2016hml} & 0.267 & 0.291 \\
DGCLR~\cite{Ding:2015rkn} & 0.262 & 0.255 \\
MVRB~\cite{Mojica:2017tvh} & - & 0.306 \\
\hline \hline
\end{tabular}
\caption{\label{tab:charmDC} The decay constants for the states
$\eta_{c}(1S)$ and $J/\Psi(1S)$. Note that the numerical values
contain a division by $\sqrt{2}$ for a consistent comparison
between different computations. The same is true for all
subsequent discussion.}
\end{center}
\end{table}

\begin{table}[h]
\begin{center}
\begin{tabular}{c|c|c}
\hline \hline
    & \multicolumn{2}{c}{decay constants [GeV]}  \\
\hline
 & $f_{\eta_{b}}$ & $f_{\Upsilon}$
\\
Experiment & --- & 0.506 \\
CI [This work] & 0.553 & 0.219 \\
Lattice QCD & 0.472~\cite{McNeile:2012qf} & 0.459~\cite{Colquhoun:2014ica} \\
BK~\cite{Blank:2011ha} & 0.501 & 0.486 \\
KGH1~\cite{Krassnigg:2016hml} & 0.547 & 0.543 \\
KGH2~\cite{Krassnigg:2016hml} & 0.535 & 0.500 \\
DGCLR~\cite{Ding:2015rkn} & 0.543 & 0.471 \\
\hline \hline
\end{tabular}
\caption{\label{tab:bottomDC} The decay constants for the states
$\eta_{b}(1S)$ and $\Upsilon(1S)$.}
\end{center}
\end{table}

The decay constants for the $\eta_{c}$ and $J/\Psi$ channels,
computed in Ref.~\cite{Bedolla:2015mpa}, are displayed in
\tab{tab:charmDC}, while the ones for the $\eta_{b}$ and
$\Upsilon$ channels, calculated in the present work, are tabulated
in \tab{tab:bottomDC}. Our results are in reasonably good
agreement for charmonia with about 20\% deviation. There is no
experimental value available for $f_{\eta_{b}}$. However, our
result for $f_{\Upsilon}$ does not fare very favorably against the
one obtained with the most sophisticated analysis of the SDEs to
date~\cite{Ding:2015rkn}, with the deviation stretching up to
around 50\%.

As discussed at length in Ref.~\cite{Bedolla:2015mpa}, we recall
that the decay constant is influenced by the high momentum tail of
the dressed-quark propagator and the
BSAs~\cite{Bhagwat:2004hn,Maris:1997tm,Maris:1999nt}, which probe
the position space wave function of quarkonia at the origin.
However, being momentum-independent, the CI results for the mass
function and BSAs have no perturbative tail. This aspect is
crucial in a better prediction of the decay constants and would
need to be built into the model in some indirect manner.

Moreover, as the quark masses become higher, mesons become
increasingly pointlike in configuration space, thus weakening the
interaction strength between their constituent quarks. It is this
observation which led us to the natural extension of the CI model
to heavy quarkonia by allowing a reduction of the effective
coupling $\alpha_{IR}$, accompanied by an appropriate increase in
the ultraviolet cutoff $\Lambda_{UV}$. However, we have retained
the parameters $m_g$ and $\Lambda_{\text{IR}}$ from the light
sector. See Ref.~\cite{Bedolla:2015mpa}, and references therein,
for further discussion.


\subsection{\label{sec:runningcoupling} On the Choice of Parameters}

\begin{table}[h]
\begin{center}
\begin{tabular}{c|c|c|c|c}
 \hline
 \hline
 quark & $\alpha_{IR}$ & $\Lambda_{UV}\,[\GeV] $ &
$\alpha$ & Normalized \\
\hline
$u,d,s$ & 2.922 & 0.905 & 3.739 & 1 \\
$c$ & 0.172 & 2.4 & 1.547 & 0.414 \\
$b$ & 0.023 & 6.4 & 1.496 & 0.400 \\
 \hline
 \hline
\end{tabular}
\caption{\label{tab:rc} The parameters of coupling ${\alpha}_{IR}$
and $\Lambda_{UV}$ are tabulated for different quark flavors.
$m_{g}=0.8\,\GeV$ and $\Lambda_{\text{IR}}=0.24\,\GeV$ are the
fixed parameters. We also provide explicit mass-scale dependent
$\alpha$ given by Eq.~(\ref{eqn:running-coupling}). The last
column yields its normalized value such that $\alpha=1$ for the
light quarks.}
\end{center}
\end{table}

As mentioned previously, the only two parameters which are varied
with the increasing quark masses are $\alpha_{IR}$ and
$\Lambda_{UV}$. For the charmonia and bottomonia, these are
determined from a best-fit to the mass and decay constant of the
pseudoscalar channel. Those for the light-quarks sector were first
obtained in Ref.~\cite{GutierrezGuerrero:2010md,Chen:2012qr}. The
arguments detailed earlier lead us to decrease $\alpha_{IR}$ with
the increasing quark mass. This change has necessarily to be
accompanied by a corresponding increase of $\Lambda_{UV}$ to
ensure we get the observed mass and decay constant of the
pseudoscalar meson. One may ask if there is a quantitative pattern
in the variation of $\alpha_{IR}$ and $\Lambda_{UV}$. In that
case, a choice of $\Lambda_{UV}$ will guide us towards the
corresponding value of $\alpha_{IR}$.

\begin{figure}[ht]
\hspace*{-0.5cm}
\includegraphics[width=0.53\textwidth]{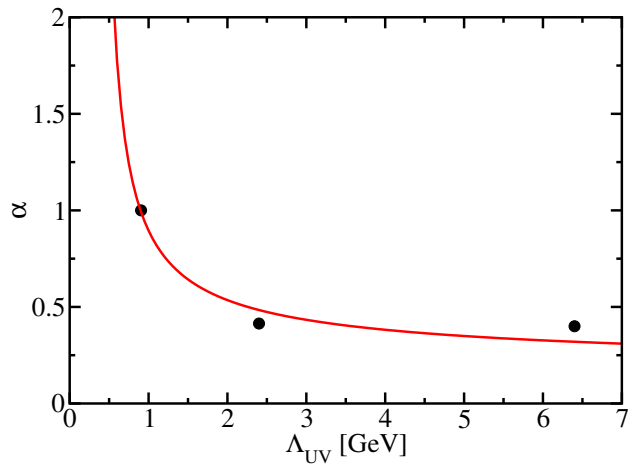}
\caption{\label{fig1} Dimensionless coupling $\alpha$ for the
contact interaction; see text. It is interesting to note that the
variation of coupling $\alpha$ as a function of $\Lambda_{UV}$ is
not far from a logarithmic decrease fitted by the function $
\alpha = a \, {\rm log}^{-1} (\Lambda_{UV}/\Lambda_0)$, where
$\Lambda_0=0.357$ GeV and $a=0.923$. }
\end{figure}


For this purpose, we define the following dimensionless coupling
%
\begin{equation}
\alpha= \alpha(\Lambda_{UV}) \equiv
\frac{\alpha_{IR}}{m_{g}^{2}}\Lambda_{UV}^2 \,.
\label{eqn:running-coupling}
\end{equation}
\noindent The advantage of this definition is that $\alpha$
naturally varies with the mass scale $\Lambda_{UV}$. This
variation is plotted in \fig{fig1}. The decrease of $\alpha$, with
respect to its value in the light-quarks sector, can be read off
from the last column of \tab{tab:rc}. Indeed, $\alpha$ is reduced
by a factor of $2.1-2.3$ in going from the light to the heavy
sector (charmonia and bottomonia), instead of the apparent large
factors quoted in Tables~\ref{tab:mcc_all_opt}
and~\ref{tab:mbb_all_opt}. More importantly, we find out that the
functional dependence of $\alpha(\Lambda_{UV})$ can be fitted
reasonably well with an inverse logarithmic curve, reminiscent of
the running QCD coupling with the momentum scale at which it is
measured.

\section{\label{sec:eff} elastic form factors}

Despite the fact that the charge-conjugation eigenstates, such as
quarkonia, do not have an EFF, we can still couple a vector
current to their constituent quarks. We expect this exercise to
yield valuable information about the internal structure of the
state~\cite{Dudek:2007zz,Bedolla:2016jjh}. Naturally, the vector
current couples to the spin-1/2 quarks through the quark-photon
interaction vertex. We briefly discuss this crucial ingredient in
the following subsection.

\subsection{\label{sec:qpv} The Quark-photon
Vertex}

   It is of fundamental importance that quarks, whose dynamics
inside hadrons is primarily dictated by QCD, also possess
electromagnetic charge. The coupling of a photon with quarks plays
an essential role in studying the internal structure of a hadron.
Being an abelian theory, the only interaction of electromagnetic
nature is the three-point quark-photon vertex.

The quark-photon vertex $\Gamma_{\mu}$ satisfies its own SDE and
is constrained by the gauge invariance of QED through the vector
Ward-Takahashi identity (WTI). Preserving this identity, and its
$Q \rightarrow 0$ limit, is crucial to the conservation of
electromagnetic current~\cite{Maris:2000sk,Maris:1999bh}. With our
treatment of the CI, the bare vertex $\gamma_{\mu}$ is sufficient
to satisfy the WTI and ensure a unit value for the charged pion's
electromagnetic form factor at zero momentum transfer. However,
given the simplicity of the model, one can readily improve upon
it. A vertex dressed consistently with our truncation is
determined by the following inhomogeneous BSE, (see
Refs.~\cite{GutierrezGuerrero:2010md,Roberts:2011wy,
Bedolla:2016jjh} for more details):
 \bea
  \Gamma_{\mu}(Q) = \gamma_{\mu} - \frac{4}{3 m_G^2} \int
  \frac{d^4q}{(2 \pi)^4} \gamma_{\alpha} \chi_{\mu}(q+P,q)
  \gamma_{\alpha} \,, \label{Eq:quark-photon}
 \eea
where $\chi_{\mu}(q+P,q)= S(q+P)\Gamma_{\mu}(Q) S(q)$.
 Therefore, the general form of
the quark-photon vertex in the RL approximation with a CI is
\begin{equation}
\label{eqn:quark_photon_rl}
\Gamma_{\mu}(Q)=\gamma_{\mu}^{T}P_{T}(Q^{2}) +
\gamma_{\mu}^{L}P_{L}(Q^{2}) \,,
\end{equation}
\noindent where $Q_{\mu}\gamma_{\mu}^{T}=0$ and
$\gamma_{\mu}^{T}+\gamma_{\mu}^{L}=\gamma_{\mu}$. The SDE for
$\Gamma_{\mu}(Q)$, Eq.~(\ref{Eq:quark-photon}), implies
\begin{eqnarray}
\label{eqn:FLFinal}
P_{L}(Q^{2})&=& 1 \,, \\
\label{eqn:FTFinal} P_{T}(Q^{2})&=& \frac{1}{1-K_{V}(Q^{2})} \,,
\end{eqnarray}
as can be consulted in Ref.~\cite{Roberts:2011wy,
Bedolla:2016jjh}. $K_{V}(Q^{2})$ is the BSE kernel in the vector
channel, \eqn{eqn:vbsagral}, within the present truncation
scheme~\cite{Bedolla:2015mpa}. Therefore, because of the dressing
of the quark-photon vertex through $P_{T}(Q^{2})$, our results for
the form factors (EFF and TFF) will develop a pole at
$Q^{2}=-m_V^2$, where $m_{1^{V}}$ is the vector meson mass. Note
that $P_T(Q^2=0)=1$ and $P_T(Q^2 \rightarrow \infty) \rightarrow
1$, leaving us with $\gamma_{\mu}$, as expected, the latter being
the statement that a dressed-quark becomes a current quark to a
large-$Q^2$ probe.

\subsection{\label{sec:pseff} $\eta_{c}$ and $\eta_{b}$ Elastic Form Factors}

A pseudoscalar meson possesses just one  vector current form factor
$F_{\eta_{c(b)}}(Q^{2})$, defined by the $\eta_{c(b)}\gamma^{*}$ vertex
\begin{equation}
\label{eqn:Psff}
\Lambda^{\eta_{c(b)}\gamma^{*}}_{\mu}(P_{i},P_{f};Q)=F_{\eta_{c(b)}}(Q^{2})(P_{i}+P_{f})_{\mu},
\end{equation}
\noindent where $Q=P_{f}-P_{i}$ is the virtual photon momentum and
$P_{i}\,(P_{f})$ is that of the incoming (outgoing) meson. In our
approach, the impulse approximation for the
$\eta_{c(b)}\gamma^{*}$ vertex reads
\begin{multline}
\label{eqn:Ps-photon-vertex}
\Lambda^{\eta_{c(b)}\gamma^{*}}_{\mu}(P,Q)=2N_{c}\int\df{k}
\Tr\left[ i\Gamma_{\eta_{c(b)}}(-P_{f})S(k_{2})\right.\\
\left. i \Gamma_{\mu}(Q)S(k_{1}) i
\Gamma_{\eta_{c(b)}}(P_{i})S(k)\right],
\end{multline}
\noindent where $\Gamma_{\mu}(Q)$ is the corresponding
quark-photon vertex. Note that $P_f = P_i + Q$. We choose
$P_{i}=P-Q/2$. Thus $P_{f}=P+Q/2$, $k_{1}=k+P-Q/2$, and
$k_{2}=k+P+Q/2$. Since the scattering is elastic,
$P_{i}^{2}=P_{f}^{2}=-m_{\eta_{c(b)}}^{2}$, which in turns implies
$P\cdot Q=0$ and $P^{2}+Q^{2}/4=-m_{\eta_{c(b)}}^{2}$, where
$m_{\eta_{c(b)}}$ is the $\eta_{c(b)}$ mass.

\begin{figure}[ht]
\hspace*{-0.5cm}
\includegraphics[width=0.53\textwidth]{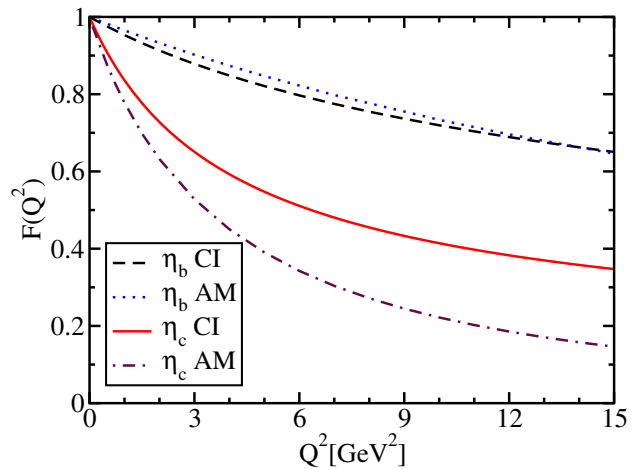}
\caption{\label{fig2} We plot CI and AM results for the $\eta_{c}$
and $\eta_{b}$ EFFs. See Ref.~\cite{Bedolla:2016jjh} for a
comparison between the CI result for the $\eta_{c}$ EFF and that
of lattice QCD from Ref.~\cite{Dudek:2007zz}. }
\end{figure}

In \fig{fig2}, we display the $\eta_{c}$ and $\eta_{b}$ EFFs,
calculated by using the dressed quark-photon vertex of
\eqn{eqn:quark_photon_rl}. For heavy mesons, such as the
$\eta_{c}$ and the $\eta_{b}$, the dressing of the vertex has a
negligible effect on the EFF; that is, in the CI, the bare vertex
trivially satisfies the WTI. However, although the time-like
sector has not been displayed in~\fig{fig2}, the $\eta_{c(b)}$
form factor has a pole at $Q^{2}=-m_V^{2}$, the mass of the
corresponding vector bound state; see Tables~\ref{tab:mcc_all_opt}
and~\ref{tab:mbb_all_opt}. This is a consequence of appropriately
dressing the quark-photon vertex. For heavy quarks, the
vector-meson pole lies deep on the time-like axis and hence hardly
affects the form factors for the space-like $Q^{2}$. Wherever a
comparison is possible, our results are harder than the ones
predicted by the Lattice QCD~\cite{Bedolla:2016jjh}; see
Ref.~\cite{Bedolla:2016jjh} for a detailed discussion.

Both for $\eta_{c}$ and $\eta_{b}$, the EFFs cease to depend on
$Q^2$ for its sufficiently large values. This is the
characteristic behavior of a point-like particle. In fact, it is
already known that the asymptotic form factors obtained with the
CI are harder than those predicted by
QCD~\cite{GutierrezGuerrero:2010md,Roberts:2010rn,Roberts:2011wy,
Bedolla:2016jjh}. This is an inevitable consequence of the
momentum independence of the
interaction,~\eqn{eqn:contact_interaction}.
As compared to $\eta_{b}$, the $\eta_{c}$ EFF increases more
steeply for $Q^{2}<0$ since the pole (the $J/\Psi$ pole for the
$\eta_{c}$ EFF) associated with the dressing of the
$c$-quark--photon vertex lies closer to $Q^{2}=0$, naturally
influencing the value of the charge radius defined as:
\begin{equation}
\label{eqn:FFradius} r_{\eta_{c(b)}}^{2}=
-6\left.\frac{\mathrm{d}F_{\eta_{c(b)}}(Q^{2})}{\mathrm{d}Q^{2}}\right|_{Q^{2}=0}
\,.
\end{equation}
\noindent Numerical values are presented in
\tab{tab:etachargeradius}. The $\eta_{c}$ charge radius compares
well with the prediction from Lattice QCD~\cite{Dudek:2007zz}. The
charge radii of $\eta_{c}$ and $\eta_b$ are also in reasonably
good agreement with SDE results with more sophisticated
Maris-Tandy model interaction~\cite{Bhagwat:2006pu,Bhagwat:2006xi}
and with a light-front quantized Hamiltonian
model~\cite{Li:2017mlw}. Clearly $r_{\eta_{b}}<r_{\eta_{c}}$,
i.e., the heavier the meson, the closer it is to being a point
particle.

\begin{table}[h]
\begin{center}
\begin{tabular}{c|c|c|c}
\hline \hline
   \multicolumn{3}{c} {\hspace{2cm} charge radius (fm) } \\
\hline  & SDE & Lattice QCD & CI
\\
\hline $\eta_{c}$ &  0.219~\cite{Bhagwat:2006pu} &
0.25~\cite{Dudek:2006ej,Dudek:2007zz}
& 0.25~\cite{Bedolla:2016jjh} \\
$\eta_{b}$ & 0.086~\cite{Bhagwat:2006xi} & --- & 0.109 [This work] \\
 \hline \hline
\end{tabular}
\caption{\label{tab:etachargeradius} The charge radius for the
state $\eta_{c(b)}(1S)$ with the CI, compared to some other
computations.}
\end{center}
\end{table}

\subsection{\label{sec:pseff} $J/\Psi$ and $\Upsilon$ elastic form factors}

The electromagnetic structure of a spin-1 meson, like $J/\Psi$ and
$\Upsilon$, is characterised by three vector current form
factors~\cite{Arnold:1979cg}. We follow Ref.~\cite{Bhagwat:2006pu}
in defining them.
\begin{equation}
\label{eqn:jpsifff}
\Lambda^{V\gamma^{*}}_{\lambda\mu\nu}(P,Q)=
\sum\limits_{i=1}^{3}T_{\lambda\mu\nu}^{i}(P,Q)F_{i}(Q^{2}),
\end{equation}
\noindent where $V=J/\Psi,\Upsilon$ and $F_{i}$, $i=1,2,3$, are
the EFFs. The kinematics remain identical to the case of
pseudo-scalar particles. However, the masses now refer to those of
$M_{J/\Psi,\Upsilon}$. Defining the transverse projector
 \begin{eqnarray}
 \mathcal{P}^{T}_{\mu\nu}(P)=
 \delta_{\mu\nu}-\frac{P_{\mu}P_{\nu}}{P^{2}} \,,
 \end{eqnarray}
  the tensors in
\eqn{eqn:jpsifff} are~\cite{Bhagwat:2006pu}:
\begin{eqnarray}
\label{eqn:tensors}
T_{\lambda\mu\nu}^{1}&=& 2P_{\lambda}\mathcal{P}^{T}_{\mu\alpha}(P^{i})
\mathcal{P}^{T}_{\alpha\nu}(P^{f}) \,, \\
T_{\lambda\mu\nu}^{2}&=&
\left[Q_{\mu}-P^{i}_{\mu}\frac{Q^{2}}{2m_{V}^{2}}\right]
\mathcal{P}^{T}_{\lambda\nu}(P^{f}) \nonumber \\
&-&\left[Q_{\nu}+P^{f}_{\nu}\frac{Q^{2}}{2m_{V}^{2}}\right]
\mathcal{P}^{T}_{\lambda\mu}(P^{i}) \,, \\
T_{\lambda\mu\nu}^{3}&=&\frac{P_{\lambda}}{m_{V}^{2}}
\left[Q_{\mu}-P^{i}_{\mu}\frac{Q^{2}}{2m_{V}^{2}}\right]
\left[Q_{\nu}+P^{f}_{\nu}\frac{Q^{2}}{2m_{V}^{2}}\right] \,.
\end{eqnarray}
A symmetry-preserving regularization scheme is crucial so that the
following WTIs are preserved~\cite{Roberts:2011wy}
\begin{eqnarray}
\label{eqn:emcurrentcons}
Q_{\lambda}\Lambda^{V\gamma^{*}}_{\lambda\mu\nu}&=&0 \,, \\
P^{i}_{\mu}\Lambda^{V\gamma^{*}}_{\lambda\mu\nu}&=&0 \, = \,
P^{f}_{\nu}\Lambda^{V\gamma^{*}}_{\lambda\mu\nu} \,.
\end{eqnarray}
\noindent The first equation follows from current conservation and
the latter two simply reflect the fact that the massive vector
meson is transverse.

A more convenient and physically relevant set~\cite{Arnold:1979cg}
of form factors, known as the charge, magnetic dipole, and
electric quadrupole form factors is given by
\begin{eqnarray}
\label{eqn:jpsigff}
G_{E}(Q^{2})&=&F_{1}(Q^{2})+\frac{2}{3}\eta G_{Q}(Q^{2}) \,, \\
G_{M}(Q^{2})&=&-F_{2}(Q^{2}) \,, \\
G_{Q}(Q^{2})&=&F_{1}(Q^{2})+F_{2}(Q^{2})+(1+\eta)F_{3}(Q^{2}) \,,
\end{eqnarray}
\noindent where $\eta=Q^{2}/(4m_V^{2})$. In the limit $Q^{2}\to
0$, these form factors define the charge, magnetic dipole and
electric quadrupole moments of the vector meson under
consideration:
\begin{eqnarray}
\label{eqn:jpsigff0}
G_{E}(Q^{2}=0)&=& 1 \,, \\
G_{M}(Q^{2}=0)&=&\mu \,, \\
G_{Q}(Q^{2}=0)&=&\mathcal{Q} \,.
\end{eqnarray}
\noindent For a point-like vector particle, the magnetic and
quadrupole moments are $\mu=2$ in units of $e/(2m_V)$ and
$\mathcal{Q}=-1$ in units of $e/m_V^{2}$,
respectively~\cite{Brodsky:1992px}.

In the impulse approximation the $V\gamma^{*}$ vertex can be
calculated from the triangular configuration~:
\begin{multline}
\label{jpsi-photon-vertex}
\Lambda^{V\gamma^{*}}_{\lambda\mu\nu}(P,Q)=2N_{c}\int\df{k}
\Tr\left[\Gamma^{V}_{\nu}(-P^{f})S(k + P^f)\right. \\
\left. i \Gamma_{\lambda}(Q)S(k +
P^i)\Gamma^{V}_{\mu}(P^{i})S(k)\right] \,.
\end{multline}
\noindent Hence if one uses a symmetry-preserving regularisation,
$G_{E}(Q^{2}=0)= 1$~\cite{Roberts:2011wy}.

\begin{figure}[ht]
\hspace*{-0.5cm}
\includegraphics[width=0.53\textwidth]{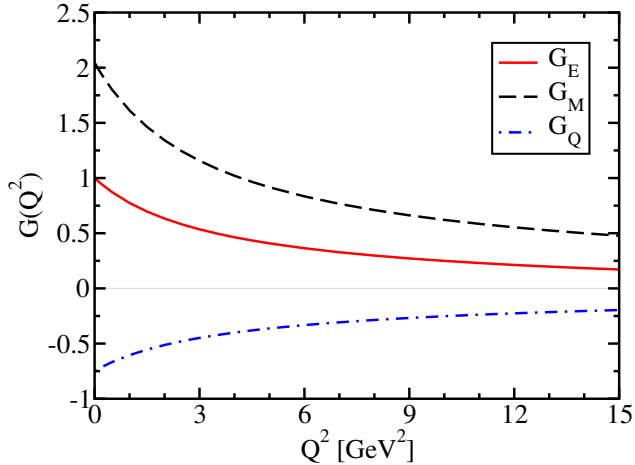}
\caption{\label{fig3}Contact interaction results for the $J/\Psi$
EFFs $G_{E}$, $G_{M}$, and $G_{Q}$.}
\end{figure}

\begin{figure}[ht]
\hspace*{-0.5cm}
\includegraphics[width=0.53\textwidth]{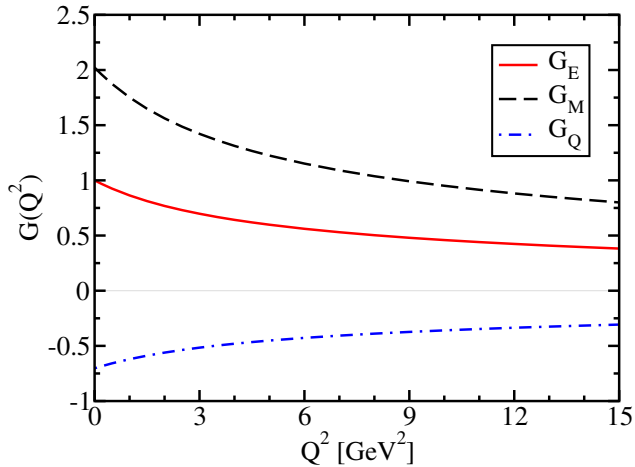}
\caption{\label{fig4} Contact interaction plots for the $\Upsilon$
EFFs $G_{E}$, $G_{M}$, and $G_{Q}$.}
\end{figure}

In Figures~\ref{fig3} and~\ref{fig4}, we plot our results for
$G_{E}$, $G_{M}$, and  $G_{Q}$ form factors for the vector mesons
$J/\Psi$ and $\Upsilon$, respectively. As these form factors
follow the same pattern as obtained for the $\rho$ meson form
factors in Ref.~\cite{Roberts:2011wy}, the same analysis applies
here. We would like to mention that as compared to the
pseudoscalars, these form factors fall faster, as $1/Q^{2}$ for
large $Q^{2}$ within the domain we explored, as opposed to the
constant value observed for the pseudoscalars (see \fig{fig2}).
This difference is due to the fact that the RL truncation with a
CI prohibits the appearance of the vector component for the BSA of
vector mesons, i.e., $F^{1^{-}}=0$ in~\eqn{eqn:vbsagral}.
Asymptotic limit of QCD predicts~\cite{Brodsky:1992px}:
 \bea
  G_{E}(Q^2):G_{M}(Q^2):G_(Q^2) \stackrel{Q^2 \rightarrow \infty}{=}
  1 - \frac{2}{3} \eta : 2 : -1 \,.
 \eea
 As noted in Ref.~\cite{Roberts:2011wy}, this relation
is recovered only for $\Lambda_{\rm UV} \rightarrow \infty$ but
spoils the convergence of integrals, implying that a vector-vector
CI cannot be regularized in a simple manner consistent with the
constraints of asymptotic QCD.

\begin{table}[h]
\begin{center}
\begin{tabular}{c|c|c|c|c|c}
\hline \hline
 & $r_{E}$ & $r_{M}$ & $r_{\mathcal{Q}}$ & $\mu$ & $\mathcal{Q}$ \\
\hline
CI  & 0.262 & 0.254 & 0.240 & 2.047 & -0.748 \\
SDE~\cite{Bhagwat:2006pu} & 0.228 & - & - & 2.13 & -0.28 \\
Lattice ~\cite{Dudek:2007zz,Dudek:2006ej} & 0.257 & - & - & 2.10 & -0.23 \\
 \hline \hline
\end{tabular}
\caption{\label{tab:jpsigffs0} Form factor radii (in fm) as well
as magnetic and quadrupole moments for the $J/\Psi$ meson.}
\end{center}
\end{table}

\begin{table}[h]
\begin{center}
\begin{tabular}{c|c|c|c|c|c}
\hline \hline
 & $r_{E}$ & $r_{M}$ & $r_{\mathcal{Q}}$ & $\mu$ & $\mathcal{Q}$ \\
\hline
CI  & 0.197 & 0.195 & 0.182 & 2.012 & -0.704 \\
SDE  & - & - & - & - & - \\
Lattice & - & - & - & - & - \\
 \hline \hline
\end{tabular}
\caption{\label{tab:upsilongffs0} Form factor radii (in fm) as
well as magnetic and quadrupole moments for the $\Upsilon$ meson.}
\end{center}
\end{table}

In Tables~\ref{tab:jpsigffs0} and~\ref{tab:upsilongffs0}, we
present charge, magnetization and quadrupole radii, as well as
magnetic, and quadrupole moments for the $J/\Psi$ and $\Upsilon$,
respectively. Reiterating the fact that for a structureless spin-1
particle, $\mu=2$ and $\mathcal{Q}=-1$~\cite{Brodsky:1992px} any
deviations from these values point to the dynamics of the internal
structure. From the tabulated results, we can infer that the
heavy-mesons produced by the CI are nearly point-like, as they
should be. Wherever comparison is possible, our results of
Table~\ref{tab:jpsigffs0} are in very good agreement with those of
SDEs and Lattice.

\section{\label{sec:tff} $\gamma\gamma^{*}\rightarrow\eta_{c(b)}$ transition
form factor}

\begin{figure}[ht]
\hspace*{-0.5cm}
\includegraphics[width=0.53\textwidth]{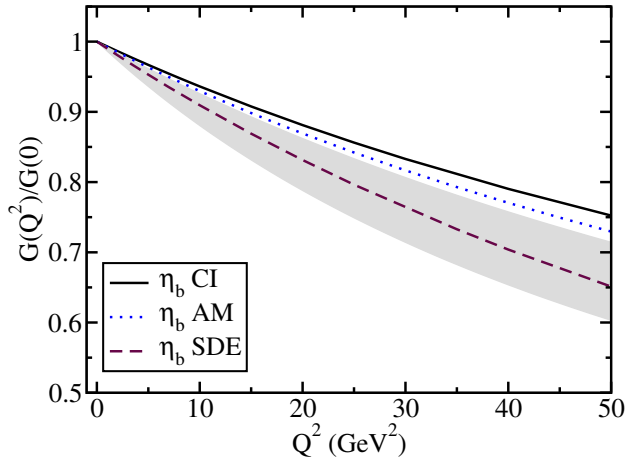}
\caption{\label{fig5} CI results for the transition form factor
for $\gamma\gamma^{*}\rightarrow \eta_{b}$ is represented by the
(black) solid line. Blue (dotted) line stems from the the AM,
\eqn{Eq:Algebraic-Model}. (Grey) band corresponds to the NNLO
nrQCD result taken from Ref.~\cite{Feng:2015uha} (the band width
expresses the sensitivity to the factorization scale). (Purple)
dashed line is the full QCD calculation of
Ref.~\cite{Raya:2016yuj}. As expected, AM does better than the CI
in its connection with the QCD predictions, though the difference
is not so conspicuous owing to the large mass of $\eta_b$.}
\end{figure}

\begin{figure}[ht]
\includegraphics[width=0.53\textwidth]{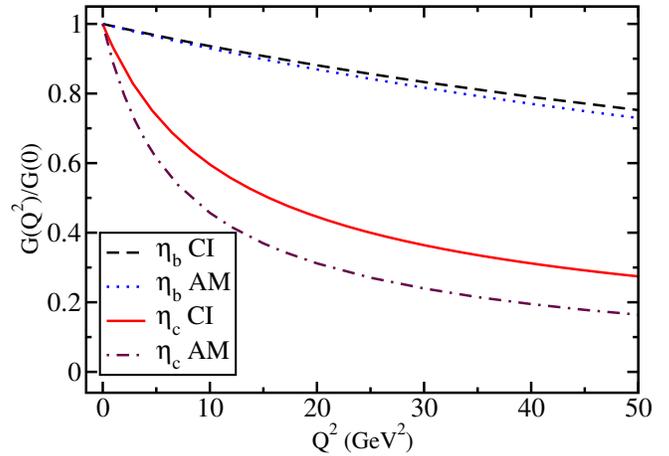}
\caption{\label{fig:quarkoniumQ2tff} Numerical results for
$G(Q^2)$ for the transition $\gamma^* \gamma \rightarrow
\eta_{c,b}$, normalized to its value at $Q^2=0$. We compare the
results for the CI and the AM. (Red) solid line is for the
$\eta_c$ in the CI, compared to the (purple) dot-dashed line
produced by the AM~\cite{Bedolla:2016jjh}. (Black) dashed curve is
the CI prediction for the CI, whereas the (blue) dotted line is
the AM result. The distinction between each pair of comparative
curves reduces substantially as the meson gets heavier. }
\end{figure}

The interaction vertex describing the
$\gamma\gamma^{*}\rightarrow\eta_{c(b)}$ transition can be
parametrized by just one form factor
$G_{\gamma\gamma^{*}\eta_{c(b)}}(Q_1^{2},Q_2^{2})$, which can be
computed from
\begin{equation}
\label{Pstransitionfull}
\mathcal{T}_{\mu\nu}(Q_{1},Q_{2})=T_{\mu\nu}(Q_{1},Q_{2}) + T_{\nu\mu}(Q_{2},Q_{1}),
\end{equation}
\noindent where $Q_{1}$ and $Q_{2}$ are incoming photon
momenta, $P=Q_{1}+Q_{2}$ is the pseudoscalar's momentum, and the
matrix element $T_{\mu\nu}(Q_{1},Q_{2})$ is given by
\begin{eqnarray}
\label{etactransition}
T_{\mu\nu}(Q_{1},Q_{2})&=&\frac{\alpha_{\text{em}}}{\pi f_{\eta_{c(b)}}}
\epsilon_{\mu\nu\alpha\beta}Q_{1\alpha}Q_{2\beta}
G_{\gamma\gamma^{*}\eta_{c(b)}}(Q_{1}^{2},Q_{2}^{2}) \nonumber \\
&&\hspace{-2.5cm} = \Tr\int\df{k}S(k_{1})\Gamma_{\eta_{c(b)}}(P)S(k_{2})
i \Gamma_{\mu}(Q_{2})S(k_{3}) i\Gamma_{\nu}(Q_{1}), \nonumber \\
\end{eqnarray}
\noindent where $k_{1}=k-Q_{1}$, $k_{2}=k+Q_{2}$, $k_{3}=k$, and
$\alpha_{\text{em}}={e^{2}}/{(4\pi)}$. In order to conform with
the experimental set up, the kinematic constraints are
$Q_{1}^{2}=Q^{2}$ and $Q_{2}^{2}=0$. We thus have $Q_{1}\cdot
Q_{2}=-(m_{\eta_{c(b)}}^{2}+Q^{2})/2$, where
$P^{2}=-m_{\eta_{c(b)}}^{2}$.

\begin{table}[h]
\begin{center}
\begin{tabular}{c|c|c|c|c}
\hline \hline
 \multicolumn{4}{c} {\hspace{2cm} interaction radius [fm] } \\
\hline & Experiment & Lattice QCD
& SDE & CI \\
$\eta_{c}$ & 0.166~\cite{Druzhinin:2010zza} & 0.141~\cite{Dudek:2007zz}
& 0.16~\cite{Raya:2016yuj} & 0.133~\cite{Bedolla:2016jjh} \\
$\eta_{b}$ & --- & --- & 0.041~\cite{Raya:2016yuj} & 0.040 [This work]\\
 \hline \hline
\end{tabular}
\caption{\label{tab:etactchargeradius} Interaction radius of the transition
$\gamma\gamma^{*}\rightarrow\eta_{c(b)}$ form factor as defined in \eqn{eqn:FFradius}.
The experimental and lattice QCD results were extracted from their respective
parametrization of data.}
\end{center}
\end{table}

In \fig{fig5}, we present the CI results for the
$\gamma\gamma^{*}\rightarrow\eta_{c(b)}$ TFF. Although not shown
in \fig{fig5}, both TFFs have a pole at $Q^{2}=-m_V^2$, as
expected. For the $\eta_{c}$, the results compare fairly well with
the {\it BABAR} data and Lattice QCD for low $Q^{2}$; see
Ref.~\cite{Bedolla:2016jjh} for a more detailed analysis. However,
for intermediate to large $Q^{2}$, the CI provides a harder form
factor, and the correct asymptotic $Q^{2}$ behavior is not
captured; see also \fig{fig:quarkoniumQ2tff}. We have neither
experimental nor Lattice QCD data for the $\eta_{b}$ TFF and
interaction radius. However, we expect similar comparisons. Due to
the same reason, in \fig{fig5}, we could only provide the
comparison of our findings with the NNLO nrQCD
result~\cite{Feng:2015uha} and a recent QCD based computation in
the SDE-BSE formalism~\cite{Raya:2016yuj}. Expectedly and
encouragingly, our plots are in the same ballpark despite the
simplicity of the model.

On the other hand, in \fig{fig:quarkoniumQ2tff}, we present our
results for $G_{\gamma\gamma^{*}\eta_{c(b)}}$. From this figure
and \fig{fig5}, it can be deciphered that both TFFs tend to a
constant value for large $Q^{2}$. However, the $\eta_{b}$ TFF
reaches this limit faster. The interaction radius of the
$\eta_{c}$ TFF, defined in~\eqn{eqn:FFradius} and tabulated
in~\tab{tab:etactchargeradius}, compares well with Lattice QCD and
the {\it BABAR} findings, as it probes the slope of the TFF for
$Q^2 \rightarrow 0$. It is also in good agreement with a recent,
more sophisticated, SDE result~\cite{Raya:2016yuj}. As with the
$\eta_{c}$'s interaction radius, we expect to achieve similar
agreement for the $\gamma\gamma^{*}\rightarrow\eta_{b}$
transition. Indeed our, result for $r_{\eta_{b}}$ agrees nicely
with that produced recently in Ref.~\cite{Raya:2016yuj}. We hope
to have corresponding Lattice QCD and experimental input at some
point in future.



\section{\label{sec:conclusions} Conclusions}

We compute the ground state spin-0 and spin-1 heavy quarkonia
masses and decay constants by using a rainbow-ladder truncation of
the simultaneous set of SDE and BSE within a CI model of QCD,
developed initially for the light quarks
sector~\cite{GutierrezGuerrero:2010md,Roberts:2010rn,
Chen:2012qr,Roberts:2011cf,Roberts:2011wy}.

It was realized in Ref.~\cite{Bedolla:2015mpa,Bedolla:2016jjh}
that the extension of the CI model to charmonia requires a
reduction of the interaction strength and a corresponding increase
in the the ultraviolet cut-off. Present article is a
generalization of this work to include bottomonia.

We find that the masses of the ground state heavy quarkonia are in
good agreement with experimental results
available~\cite{0954-3899-37-7A-075021} as well as SDE
calculations with QCD based refined
truncations~\cite{Blank:2011ha,Ding:2015rkn}. The decay constants,
however, are a bit lower. We have also computed the EFFs for
$\eta_{c}$, $\eta_{b}$, $J/\Psi$, and $\Upsilon$, the transition
form factors of the $\eta_{c}$ and $\eta_{b}$ and the
corresponding radii.

The dressing of the heavy-quark--photon vertex, consistent with
the model truncation and the WTI, ensures that the form factors
posses a vector meson pole at $Q^{2}=-m_V^{2}$. However, since the
vector meson mass is large, the effect of the meson vector pole on
the charge radii is negligible. Our form factors, however, have
better quantitative agreement with data and/or other calculations
for small values of $Q^{2}$, whenever available. They are harder
for intermediate and large $Q^2$. For the pseudoscalars, the both
form factors tend to a constant for $Q^{2}\rightarrow\infty$,
which is a consequence of the momentum-independent interaction. On
the other hand, the vector component of the vector meson is zero
and the corresponding EFFs tend to behave as $1/Q^2$ for large
$Q^{2}$.

Our results for charge radii are in very good agreement with
Lattice QCD data and experimental results, whenever available;
furthermore comparison between the charga radii for charmonia and
bottomonia confirms that the heavier the meson the closer it is to
being a point particle, justifying our reduction of the strength
of the coupling.

All this is an encouraging first step towards a comprehensive
study of heavy hadrons in this approach. Immediate next steps will
involve flavored mesons and baryons. Our goal is to provide a
unified description of light and heavy hadrons within the CI
model.

\section{Acknowledgments}

The authors acknowledge financial support from CONACyT, M\'exico
(doctoral scholarship for M.A.~Bedolla; postdoctoral Contract
No.~290917-UMSNH for J.J.~Cobos-Mart\'{\i}nez and research Grant
No. CB-2014-242117 for A.~Bashir). This work has also partly been
financed by the CIC-UMSNH Grant 4.10. J.J.~Cobos-Mart\'{\i}nez
also acknowledges the support of Conselho Nacional de
Desenvolvimento Cientifico e Tecnologico (CNPq,Brazil), Grant No.
152348/2016-6 and  Instituto Nacional de Ciencia e Tecnologia -
Fisica Nuclear e Aplica\c c\~oes (INCT-FNA, Brazil) CAPES:
88887.145710/2017-00.


\bibliography{SDEBSEReferences}

\end{document}